\documentclass[12pt]{article}
\usepackage{fullpage,isolatin1,times,verbatim}
\usepackage{psfig}

\title{Towards Solving the Interdisciplinary \\ Language Barrier Problem}
\author{Sébastien Paquet}

\begin{document}

\bibliographystyle{plain}

\maketitle

\vspace*{3.0cm}

\emph{There are two great secrets to success
      in life. The first is to not tell
      everything you know.}

Anonymous
\vspace*{1.0cm}

\emph{Everybody laughs in the same language. }

Anonymous

\subsection{Abstract}

This work aims to make it easier for a specialist in one field to find
and explore ideas from another field which may be useful in solving a
new problem arising in his practice. It presents a methodology which
serves to represent the relationships that exist between concepts,
problems, and solution patterns from different fields of human
activity in the form of a graph. Our approach is based upon
generalization and specialization relationships and problem
solving. It is simple enough to be understood quite easily, and
general enough to enable coherent integration of concepts and problems
from virtually any field. We have built an implementation which uses
the World Wide Web as a support to allow navigation between graph
nodes and collaborative development of the graph.

\section{Introduction}

\subsection{The Ills of Specialization}

The problem we are concerned with in this work is extremely common in
today's world. Anyone who has consulted medical doctors without having
himself or herself a good grasp of medicine has encountered a language
barrier first-hand : whereas the doctor (usually) knows what
he\footnote{This text uses the masculine to include the feminine.} is
doing and will be able to discuss it with other experts in the same
field, his patient will understand very little of what is going on in
such a dialogue, because it will be replete with terms that mean
nothing to him. Moreover, frequently the doctor will not be able to
fully convey his understanding to the patient. This can be a
frustrating predicament, especially as the issues under discussion may
greatly matter to patient.

To \emph{specialize} is to concentrate on a particular, necessarily
restricted activity or field of study. Observing the contemporary
world, it is not difficult to discern a massive trend towards the
specialization of humans, which translates into an explosive growth in
the number of fields of human activity. Specialization is a way of
increasing overall efficiency; one of the better known examples is
Henry Ford's introduction of the assembly line for manufacturing
cars. By assigning to each member of the production team a single task
to be performed on the cars as they passed by slowly down the assembly
line, Ford realized a tremendous gain in efficiency compared to
traditional assembly methods.

While specialization is desirable from that particular point of view,
it also has a downside: the more a person becomes specialized, the
less he can meaningfully discuss problems that matter to him with
other people. In effect, the specialist is very often restricted to
collaboration with other specialists in the same area. As one gets
more and more specialized, pools of colleagues grow ever smaller. This
is unfortunate because discussions with specialists in another area
often prove to be fertile ground, as ideas and strategies which were
first developed in one field often turn out to be adaptable to a
problem in a different field.  Indeed, breakthroughs often result from
interdisciplinary collaboration: it is not unusual that effective
tools for tackling a long-standing problem in an area are found in
another area. For instance, the recent scientific successes of genome
sequencing owe much to the collaboration between biologists and
computer scientists. Another adverse consequence of the isolation
arising from specialization is that many people reinvent the wheel for
themselves because they are unaware of similar work that has been done 
elsewhere.

Interdisciplinary communication is thus desirable from the point of
view of progress; that is, it is helpful in solving problems,
especially the more important and challenging ones. Consequently,
finding efficient ways of communicating with outsiders is becoming an
increasingly pressing problem for people who are not content with
speaking only with an inner circle of colleagues. A way to alleviate
the problem is to learn another specialty, which will provide
opportunities for discussion with a larger circle of people. However,
learning a specialty usually involves a considerable time investment;
moreover, there are so many different specialties\footnote{Crane and
Small reported that there were 8530 definable knowledge fields in
1987. It is reasonable to presume that the figure has grown
significantly since then.}  that even selecting a promising one can be
a difficult problem in itself.

\subsection{A Language Issue}

Human communication is an activity whose goal is to convey meaning
from one person to another. Language is a set of signals (such as
words, images or gestures) which serves to conduct it. In order for
two people to communicate, it is necessary that they agree on signals
and their meanings. A language barrier exists whenever the signals one
person uses are not recognized or have different meaning for the other
person.

Since different specialties are concerned with different things,
specialists develop different languages that enable them to
communicate efficiently amongst themselves. The advantage in
developing a specialized language is that of conciseness. Expressing a
complex, multi-faceted construct in a single word is preferable to
always using everyday (commonly used) language. This is because the
unequivocal description of the construct in everyday language is
usually much longer. As a case in point, consider the fact that
mathematical equations used to be expressed in words. Following the
development of a symbol system providing shorthand representations for
recurring expressions (i.e. \emph{x} being substituted for the phrase
``the unknown''), it became easier to think about complicated
mathematical problems. However, at the same time an additional
requirement was imposed unto whoever wanted to understand what
mathematicians were doing: it became necessary to learn their 
specialized language.

In the time of Leonardo da Vinci, it was possible for a dedicated
individual to become reasonably versed in most existing disciplines,
providing ample occasion for cross-fertilization between
fields. Sadly, nowadays this is no longer possible. The
interdisciplinary language barrier problem is universal, because one
can be trained in no more than a few specialties, and consequently
cannot easily understand the languages that are used in the other
specialties.

\subsection{Connections Between Languages}

How does one convey the meaning of words\footnote{in what follows, we
use ``word'' to mean any kind of signal.} belonging to a language
which they understand to someone who has never heard of them before?
(This is a central problem in teaching.) A well-known strategy
consists in using words known to both teacher and student to explain
what a new words means. Those will either be words taken from everyday
language, or specialized words which are known to be understood
because an area of overlap exists between the teacher's and the
student's backgrounds.

We wish to stress the point that this overlap, the common ground
between teacher and student, can be a major determinant of the
efficiency of this process. For instance, if both master the language
of mathematics at a sufficiently advanced level, the teaching of
elementary physics should be greatly facilitated. This is because
numerous and deep connections have been established between those two
fields. Thus a learner acquainted with the concept of derivative will
most likely quickly understand that of velocity, seeing it as a
special case of a derivative of a spatial variable with respect to a
temporal variable.  This \emph{specialization relationship} between
concepts can be seen as a
\emph{shortcut} from one field to the other, since understanding one of
the concepts enables one to understand the other quickly, \emph{without
building it from the ground up}. A similar argument could be made
concerning the situation of someone familiar with the concept of
velocity who wishes to understand that of derivative. If we abstract
out the labels of space and time from the concept of velocity we get
that of derivative. In that direction, the shortcut is rather a
\emph{generalization relationship}. Specialization and generalizations 
are two sides of the same coin: the former consists in putting
particular things in as-yet-empty boxes, while the latter consists in
removing them.

The last example referred to recognizing something new as a particular
form of something that is already understood, and vice
versa. Sometimes connections between concepts are not as obvious. For instance, a
physicist who is introduced to the concept of chemical reaction speed
might perceive a similarity between it and the concept of velocity,
which might help him grasp the former concept. What happens here is
that both concepts may be seen as special cases of a common, more
general concept: that of a derivative with respect to time. In this
case the shortcut is an indirect, two-link path from the particular to
the general and back to the particular. Note that even though the
learner may perceive the similarity between two concepts, he may not
readily express the common abstraction in words.

Shortcuts of the three types mentioned above (specialization,
generalization, and similarity) have much value because they enable
learners to quickly learn about a new topic, capitalizing on their
mastery of a specialty. Having access to a large, organized body
of such shortcuts, would provide specialists with a quick and rather
painless way of learning about new topics. It would consequently
enable them to discuss meaningfully with specialists from other fields,
thereby giving them a chance to overcome the interdisciplinary
language barrier with much less effort than is currently required.

The natural question which arises from the previous discussion is the
following one: are there really that many shortcuts between the
concepts belonging to different fields of human activity?  If so, are
there ways of identifying them, organizing them, and sharing them?
The goal of the present work is to examine those questions and provide
elements of answers to them.

\subsection{Mathematics as ``Shortcut Science''?}

The previous examples, as well as simple observation of the methods
used in several sciences, point to the discipline of mathematics as a
kind of substrate underlying many specialized concepts. Indeed,
mathematics does provide a kind of integrative unity to a very large
body of knowledge comprising most of physics and chemistry, parts of
engineering, computer science and economics, etc. One could think of
it as a well-organized, useful network of passageways between those
fields. A mastery of mathematics will definitely help someone learn
any of those fields.

However, the network is far from completely connecting every concept in
every field and subfield. If such were the case, able mathematicians
would likely rule the world. What is missing? What kinds of knowledge
have not been abstracted out and organized by mathematicians?

Insofar as it can be considered as knowledge, \emph{know-how} or
\emph{problem solving knowledge}, which relates to the actions
that enable one to accomplish a particular task or to reach a
particular goal, is an obvious and glaring omission. Among other
disciplines, computer science has made a few forays into this area,
especially as regards the \emph{specification} of know-how, but a
proper and general \emph{organization} of know-how is nowhere to be
found at present. Many problem solving strategies have been
identified, a few relationships have been identified between them, but
for most problems that appear for the first time there is currently
nowhere to look for potentially appropriate strategies.

Problem solving is a central human activity. Most occupations are
defined by the set of problems they are concerned with. Problem
solving knowledge that is specific to an occupation is often referred
to as the ``tricks of the trade'', and (sadly) is seldom well
documented. Acquiring it quite often involves a sometimes costly
process of trial-and-error which is called ``experience''. More often
than not, success in an occupation depends on achieving mastery of
those tricks. The usefulness of properly documenting problem solving
processes lies, among other things, in the potential it has for
reducing the amount of trial-and-error that is necessary to learn a
trade.

\subsection{Scope of this work}

This work makes the (admittedly audacious) suggestion that a simple
approach laid upon the foundation of specialization and generalization
relationships and problem solving strategies may enable concepts,
problems and solution strategies from all fields of human endeavour
which are describable by language to be structured together in a
coherent, usable manner. This structure may be seen as a large graph
linking these entities together. Moreover, it proposes a practical
means of building and providing access to such an organized body of
knowledge which is based on currently available technology.

Of course, our conjecture cannot be proved true without exhaustively
organizing all (symbolic) human knowledge, a task which is beyond the
reach of any individual. However, demonstrating it to be false would
require identifying something which demonstrably does not relate to
anything. The mere act of defining such a thing would put it in
relationship with something, which would create a contradiction.

Having put aside questions of absolute truth or falsity regarding our
claim, one might still legitimately attempt to evaluate the
practicality of the methodology presented here.  In order to address
this question, we have applied the methodology to produce a
small-scale prototypical structure comprising concepts from
diverse fields of activity, to test the practicality and the general
applicability of our approach. 

\subsection{Citation Warning}

Texts such as this one are usually replete with references to related
work by other researchers. As the topic under discussion here is quite
broad, every reader will undoubtedly find relationships between this
work and other work, perhaps even his own. Without the shadow of a
doubt, the present work reinvents several wheels which have already
been invented in other fields of study.

However, the reader will find a remarkable paucity of citations in
this document. There are two reasons for this state of affairs, one
benign, the other more serious. The first is that the author will
never have time to read and understand everything in every discipline
which could be related to this work. The second is that, because the
problem considered here is intrinsically interdisciplinary, the
author doesn't have at his disposition proper \emph{means} of finding
everything which could be related to this work. This unfortunate
predicament actually constitutes the original motivation for this work.

\subsection{Structure of the Document}

The remainder of this document further explains and provides support
for what has been suggested in this introduction. 

Section \ref{approach} explains our approach to the problem. It is
divided in two. Subsection \ref{graphs_nodes_links} explains the
basics of graphs, which we subsequently use to explain our
approach. Other ways of describing the same approach are possible, but
the idea of a graph is intuitive and allows for easy-to-grasp visual
representations.

Subsection \ref{concepts} argues that any concept
that is expressible in a language is defined by way of its
relationships to other concepts. It describes the basic structure of
a type of graph node which may serve to represent a concept, and types
of links which may originate from such a node.

Three particular kinds of concepts are central to our work and are
explained in Subsection \ref{special_concepts}. They are: problems,
solution patterns, and strategies. Another useful concept is that of
domain, which is explained last.

Section \ref{implementation} describes our prototypical implementation of
a concrete example of a product of our approach. We conclude in
section \ref{conclusion} by summarizing this work and providing
directions for future research.

\section{A General Approach to Organizing Concepts and Problems}
\label{approach}

\subsection{Graphs, Nodes and Links}
\label{graphs_nodes_links}

\begin{figure}[htbp]
\centerline{\hbox{\psfig{figure=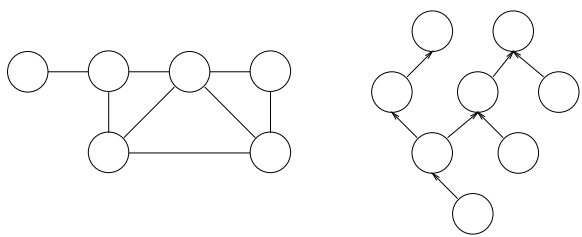,height=3.0cm}}}
\caption{Examples of graphs. Nodes are represented as circles and
links as line segments (indicating symmetrical relationships) or
arrows (indicating asymmetrical relationships).}
\label{graph}
\end{figure}

A \emph{graph} is a set of abstract entities called \emph{nodes} that
are connected by abstract entities called \emph{edges}. The word
``edge'' has a geometrical connotation which may be confusing in the
context of this work; we use the word ``link'' instead. Links
represent relationships that exist between nodal entities.  A link
which represents an asymmetrical relationship (for instance, the
relationship between parent and child) has a direction, that is, one
of its endpoints is labelled as the origin node and the other is
labelled as the destination node.  Sample visual representations of
graphs are found in figure
\ref{graph}.

The concept of a graph is useful in a multitude of contexts, for
instance for representing transportation or communication networks. It
has also been extensively used to represent relationships between
concepts. Our approach uses nodes to represent concepts, and links to
represent specific kinds of relationships that exist between them.

\subsection{Concepts}
\label{concepts}

Dictionaries define the meaning of words using other words.  Teachers
do the same with students who are learning new ideas. This suggests
the general statement that a concept that is expressible in a language
is defined by way of its relationships with other concepts. This is
not something which can be proved. The only way to go further in this
work is to temporarily accept that it may be true and see where this
leads.

We have said that we want to define a type of graph node which may
serve to represent any concept. We have already explained what we mean
by definition, generalization and specialization. Our proposition is
the following. A concept node features:

\begin{enumerate}
\item a \emph{name}, which is a signal which may be used to refer to
the concept among people who understand it \footnote{We remark that
words often have multiple meanings. This means that one particular
expression may appear as a name in a number of distinct concept
nodes. Obviously, the rest of the node contents disambiguates the name.};
\item a \emph{definition} which contains links to the concepts which
are involved in it;
\item a set of links to concepts that are \emph{generalizations} of
the concept;
\item a set of links to concepts that are \emph{specializations} of
the concept;
\item a set of links to other \emph{concepts} whose definition
directly involves the concept;
\item a set of links to \emph{problems}\footnote{We explain what we mean by a
problem in section \ref{problems}.} in which the concept is directly involved; and
\item a link to a \emph{domain}\footnote{We explain what we mean by a
domain in section \ref{domains}.} which this concept primarily relates to.
\end{enumerate}

We give an example of a concept node which could describe
the concept of an ordinary quadratic function.
\begin{enumerate}
\item The name is the expression ``real-valued quadratic function of a real 
variable''.

\item The definition is the phrase 
``A \emph{real-valued polynomial function of a real variable} involving
\emph{terms} of the \emph{second degree} at most.'' Italicized words indicate that 
links to the corresponding concept nodes exist.

\item Generalizations could include, but are not limited to:
``polynomial function solvable by radicals'', ``function having a
global extremum'', and ``high school level mathematical
concept''. Note that the first generalization classifies the concept
from the standpoint of algebra, the second one highlights a property
that could be useful in an optimization context, and the third
indicates a property that could be useful in an educational
context. The three are quite different; their relevance depends on
what one is looking for.

\item  Specializations could include: 
``real-valued quadratic function of a real variable having no real
root'', ``real-valued quadratic function of a real variable having
one real root'', ``real-valued quadratic function of a real variable
having two real roots'',
``real-valued quadratic function of time'',
``real-valued quadratic function of a spatial variable''.
Here the first three specializations have a mathematical flavour
while the latter two have a physical flavour.
\end{enumerate}

Let us recall at this point that the overall goal is to obtain a graph
that is easy for humans to use. This implies that someone who is
creating a node must strive to ensure that none of the above two lists
grow to become unmanageably large. In order to achieve this, a guiding
principle needs to be applied. Our prescription is the following:
\emph{systematically hide complexity by introducing new, meaningful nodes
wherever possible}. What this means for instance is that instead of
listing ``real-valued quadratic function of straight-line distance'',
``real-valued quadratic function of geodesic distance'', etc. as
specializations, one should instead bundle them together into
``real-valued quadratic function of a spatial variable''. The same
principle should be applied for generalizations. A similar principle
is often applied successfully in the context of writing, where
explicit structure serves to help the reader deal with the overall
complexity of the text.

\begin{enumerate} \setcounter{enumi}{4}
\item An example of a concept whose definition directly involves this
concept is that of a paraboloid.

\item Problems could include links to the following problem
nodes: ``Obtain the roots of a real-valued quadratic function of a
real variable'', ``Find the extremum of a real-valued quadratic
function of a real variable'', ``Teach the concept of quadratic
function''.

\item Finally, the domain link could connect the concept node to the domain of
mathematics.

\end{enumerate}

\subsection{Special Concepts}
\label{special_concepts}

In the last subsection, we defined the structure of nodes which
represent concepts. Here we define slightly different structures for
nodes which describe problems, solution patterns, strategies, and domains.

Subsection \ref{problems} explains what we mean by a problem. It
describes the basic structure of a type of graph node which may serve
to represent a problem, and types of links which may originate from
such a node.

Subsection \ref{solution_patterns} explains what we mean by a solution
pattern to a problem. It describes the basic structure of a type of
graph node which may serve to represent a solution pattern, and types
of links which may originate from such a node.

Subsection \ref{strategies} explains what we mean by a strategy. It
describes the basic structure of a type of graph node which may serve
to represent a strategy, and types of links which may originate from
such a node.

Subsection \ref{domains} explains what we mean by a specialized domain
of activity and argues that a domain can be approximately defined by
the set of concepts, problems and solution patterns which it is
concerned with. It describes the basic structure of a type of graph
node which may serve to represent a domain, and types of links which
may originate from such a node.

At this point in our exposition, we would suggest that the reader take
a look around our prototype implementation, which is described in
section \ref{implementation}. This should make it easier to understand
what follows.

\subsubsection{Problems}
\label{problems}

Up until now we have talked about problems without defining more
precisely what we meant by that term, hoping that the reader has
encountered enough problems in his life to have an intuitive grasp on
that concept.

Our conception of a \emph{problem} is similar to that proposed by
Polya\cite{Polya}: \emph{a goal in a context which we can express
clearly, but that is not immediately accessible}. This is a very broad
definition. Examples of common problems are: to find something to eat
in a cluttered refrigerator, to obtain recognition from others, and to
understand something. 

What constitutes a problem for a person is not necessarily a problem
for another. For instance, for most people getting out of bed is an
immediately accessible goal, while this is far from being the case for
a baby. Thus a possible line of action for someone who does not know
how to solve a problem is to collaborate with someone who does. (This
may be a problem in itself.)

A problem is fully specified when both goal and context are
unequivocally specified in terms of concepts. We propose to represent
problems in nodes which feature:

\begin{enumerate}
\item a (possibly empty) set of \emph{names} used for referring to the
problem;
\item a \emph{description} of the problem, comprising goal and context;
\item a set of links to \emph{generalizations} of the problem; that is,
instances of the problem with a generalized form of the goal, a
generalized form of the context, or both;
\item a set of links to \emph{specializations} of the problem; that is,
instances of the problem with a specialized form of the goal, a
specialized form of the context, or both;
\item a set of links to \emph{solution patterns} to more complex
problems which require, as a substep, that this problem be solved,
providing motivation for solving the problem;
\item a set of links to \emph{solution patterns} that apply to the problem;
\item a link to a \emph{domain} which this problem primarily relates to.
\end{enumerate}

We give an example of a problem node which could describe the problem
of solving an ordinary quadratic equation, given that we know the
quadratic formula $\frac{-b\pm\sqrt{b^2-4ac}}{2a}$. We ought to warn
readers who are used to solving this problem without even thinking
about it that seeing the problem from the perspective of someone who
is tackling it for the first time will likely require some effort.

\begin{enumerate}
\item A name could be ``Solve an ordinary quadratic equation''. Note 
that it is easier for a specialist to refer to the problem by its
name, while it is easier for a non-expert to refer to the problem by
its longer but more explicit description.
\item The description of the problem could be the following:
``Obtain the \emph{roots} of a \emph{real-valued quadratic function
of a real variable}, using the \emph{quadratic formula}''. Links to
the relevant concepts, i.e. that of a root, that of a real-valued
quadratic function of a real variable, and that of the quadratic
formula are provided.
\item An example of a generalization of the problem could be:
``Obtain the roots of a polynomial function solvable by radicals,
using an appropriate formula.''
\item A (perhaps a little contrived) example of a specialization of
the problem could be: ``Obtain the roots of a real-valued quadratic
function of a real variable of the form $ax^2+c$, using the quadratic
formula.''
\item An example of a solution pattern which involves solving this
problem could be that consisting in finding the intersection of a line
with a parabola by (1) finding the corresponding quadratic equation; and
(2) solving it.
\item A solution pattern that is applicable to the problem
could be that consisting in computing the solution by feeding the
polynomial coefficients into the quadratic formula.
\item Finally, the domain link could connect the problem node to the domain of
mathematics.
\end{enumerate}

\subsubsection{Solution Patterns}
\label{solution_patterns}

Solving a problem amounts to finding a precise line of action which
will enable one to attain the goal. The line of action is made up of a
series of subproblems to be solved. This is what we mean by the term
\emph{solution pattern} \footnote{The equivalent term in computer
science is \emph{algorithm}. \emph{Procedure} is another word which
refers to the same concept}. Finding (and remembering) a solution
pattern is interesting because it is reusable once found: if the same
problem arises again, it suffices to apply the same pattern again to
solve it. For instance, once the standard solution pattern to ordinary
quadratic equations has been remembered, any such equation can be
solved without hesitation. The more general the solution pattern, the
wider its applicability. Of course, a very general solution pattern
can hide much of the actual complexity because the subproblems it
features can be complex problems in themselves. By contrast, a very
specialized solution pattern is quite straightforward to put into
application, but has a correspondingly restricted applicability.

A solution pattern is fully specified when all the subproblems which
compose it are identified. The solution pattern is guaranteed to be usable
if every one of its subproblems has a known solution pattern. If such
is not the case, the solution pattern may be usable, but this is
conditional upon finding solution patterns to every subproblem.

We propose to represent solution patterns in nodes which feature:

\begin{enumerate}
\item a link to the associated \emph{problem};
\item a link to a \emph{strategy} which generalizes the solution pattern (strategies are defined in the next section);
\item a \emph{specification} of the solution pattern, containing links to the 
\emph{subproblems} that it entails;
\item a link to a \emph{domain} which this solution pattern primarily relates to.
\end{enumerate}

We give an example of a solution pattern node which could describe a
way of solving an ordinary quadratic function.
\begin{enumerate}
\item A link to the problem ``Obtain the roots of a real-valued
quadratic function of a real variable, using the quadratic formula'' could be provided.
\item A link to the strategy ``Obtain a result from an appropriate input and a procedure'' (explained next) could be provided.
\item The solution pattern consists of a single subproblem: ``Evaluate the quadratic formula on the polynomial coefficients''.
\item Finally, the domain link could connect the problem node to the domain of
mathematics.
\end{enumerate}

\subsubsection{Strategies}
\label{strategies}

This work makes the (perhaps controversial) assumption that every
problem solving process can be modelled as a sequence of two steps:
(1) putting the problem in relationship with a very general, abstract
form of problem for which the (abstract) solution is known; (2)
translating the abstract solution into terms that are specific to the
problem. We call \emph{strategy} an abstract problem form accompanied
by its associated solution which has wide enough applicability to be
used in almost any domain. A strategy captures the essential aspects
of a solution pattern. Strategies are interesting because they suggest
reusable forms of solution for wide classes of problems.

We illustrate the process of going from solution pattern to
strategy by considering the following simple problem. Suppose a friend has
secretly chosen a number between 1 and 10 and asks you to repeatedly
guess what it is. Each time you make a guess he will tell you whether
your guess is indeed the secret number. The problem is to find the
secret number. You choose a simple solution pattern: you try every
number in order, from 1 to 10, until your friend confirms that you
have found the secret number. This is an appropriate solution pattern,
in the sense that you are guaranteed to find that number.

What would be the strategy in such a case?  It is often hard to tell
what is going on in one's head; we can only make conjectures as to
what really happened. Our point of view is that strategies can be
identified, which may or may not correspond to the actual mental
process.  Although knowing more about that process would certainly be
interesting, the only goal we are concerned with here is finding an
abstract form of the problem for which a valid solution is known. The
important thing to recognize is that once we have found a potentially
appropriate strategy, it is easy to check whether is indeed qualifies
as a generalization of the solution pattern at hand.

What can be abstracted out of our problem without essentially changing
it? First, note that the concept of number is not essential, for we
could have replaced numbers with letters of the alphabet or
billiard balls. The important thing is that we had a
finite set of possible objects to choose from. Second, what is
essential in your interaction with your friend? Surely your friend
could be replaced by a machine which tells you whether you have found
the right object. The essential thing here is that you have a way of
checking whether a given object satisfies the property of being the
secret object. Finally, what was the use of counting to ten? It was to
provide a way of selecting an untested object at each guess.

Thus the abstract version of our problem is the following: ``Find an
object satisfying a particular property in a finite set of objects,
given a way to check an object for the property and a way to find an
untested object''. The strategy solution pattern is the following:
``As long as an object satisfying the property is not found, find an
untested object and test it''. People seem to use such a strategy
everyday; finding something to eat in a cluttered refrigerator,
shopping for clothes or selecting an appropriate screwdriver from a
toolbox are examples of problems where this strategy applies.

In a sense, strategies are even more useful than domain-specific
solution patterns in the sense that they are relatively easy to grasp
and apply to large sets of problems. They are also very useful from an 
interdisciplinary standpoint because they provide clear connections
between methods that are used in different fields.

We propose to represent strategies in nodes which have a structure
similar to that of problems:
\begin{enumerate}
\item a (possibly empty) set of \emph{names} used for referring to the
problem;
\item a \emph{description} of the problem, comprising goal and context;
\item a set of links to \emph{strategic generalizations} of the
problem; that is, strategies that apply to problems involving a
generalized form of the goal, a generalized form of the context, or
both;
\item a set of links to \emph{strategic specializations} of the
problem; that is, strategies that apply to problems involving a
specialized form of the goal, a specialized form of the context, or
both;
\item a \emph{specification of the solution pattern} to the problem,
containing links to the \emph{strategic subproblems} that it entails;
\item a set of links to \emph{domain-specific specializations} of the
strategy; that is, domain-specific solution patterns which are
specializations of the strategy;
\item a link to a \emph{domain} named ``strategies''.
\end{enumerate}

In the example of the applying the quadratic solution formula, the
associated strategy node could take the following form:
\begin{enumerate}
\item Possible names could include ``Apply a recipe'', ``Do it by the book''.
\item The description of the problem is ``Obtain a result from an
appropriate input and a procedure''.
\item A possible generalization could be the strategy ``Obtain a
result from a procedure and an input'', the precise nature of the input (for
instance, as being appropriate versus almost appropriate\footnote{ We
characterize an input as almost appropriate with respect to a
procedure if we have a way of converting it to an input 
that is appropriate for that procedure.})
being abstracted out.
\item A specialization could be the problem ``Effortlessly obtain a
result from an appropriate input, a procedure, and a helper''\footnote{Calling upon a
computer or a graduate student to perform the work should come to the
minds of experienced researchers.}.
\item There would be a single strategic subproblem in this case:
``Apply the procedure to the input''.
\item A domain-specific solution pattern could be the one described in 
the previous subsection;
\item Finally, the domain link would connect to the domain of strategies.
\end{enumerate}

\subsubsection{Domains}
\label{domains}

The concept of domain provides a convenient way of classifying
specialized concepts, problems and solution patterns which are somehow
related together. Almost everyone today thinks of himself as a
specialist in a particular domain. Thus we have quantum cosmologists
who are physicists who are specialized in quantum cosmology;
neurosurgeons who are medical doctors who are chiefly concerned with
brain surgery; hitmen who are criminals that specialize in murder; and
so on. Being able to label people is useful. Upon learning that
someone has expertise in a particular domain, we know that this person
masters the concepts and problems of the domain. If we need to solve a
problem which we know to be related to that domain, we know that this
person may be able to help.

We propose to represent domains in nodes which have the following
structure:
\begin{enumerate}
\item A \emph{name} which people who are specialized
in the domain use to recognize each other;
\item A set of \emph{generalizations} which are disciplines which may
be considered to be the roots of the domain; usually this is the
domain in which specialists of the domain are first trained.
\item A set of \emph{specializations} which are disciplines which may
be considered to be the branches of the domain;
\item A set of \emph{prominent concepts} which are the central
concepts used in the domain;
\item Finally, a set of \emph{prominent problems} which are the central
problems that a specialist in the domain knows how to solve.
\end{enumerate}

In the case of mathematics, the domain node could read as follows:
\begin{enumerate}
\item The name is ``mathematics''.
\item A possible generalization could be ``science''.
\item Possible specializations could include: ``fundamental
mathematics'', ``algebra'', ``number theory'', ``geometry'',
``topology'', and ``mathematical analysis''.
\item Possible prominent concepts could include: ``set'', ``number'',
``function'', ``axiom'', ``proof'', and ``definition''.
\item Possible prominent problems could include: ``prove a
proposition'' and ``find an unknown from data''.
\end{enumerate}

Domain boundaries are quite fuzzy and constantly evolving.  New
domains pop up ever more quickly as the trend towards specialization
accentuates.  On the other hand, some domains become obsolete. Thus we
see less and less blacksmiths, alchemists and carriage makers as time
goes by. An unfortunate aspect of this kind of natural selection among
specialties is that interesting solution patterns may die along with
their trade unless those processes are documented before their
complete extinction. We remark that a graph such as the one described
in this work could obviously represent the forgotten specialties
alongside the current ones.

\section{An Implementation}
\label{implementation}

The previous section described a basic architecture for organizing
concepts and methods. Here we describe a simplistic but concrete form
which this architecture may take. We have built a prototypical knowledge
base which uses this basic design. We are certain that many
improvements could be made to further facilitate navigation among
concepts.

The World Wide Web (WWW) consists of documents that are available on
the Internet. These documents (``web pages'') are written in
hypertext, which means that they may contain links to other
documents. There is no restriction on what documents a given document
may link to, as long as the target documents are themselves available
somewhere on the Internet. Users of the Web use WWW browsing software
to view documents and can move over to linked documents with a simple mouse
click.

Our implementation uses this technology in a very straightforward
manner. Each concept, problem, solution pattern or strategy is explained
on a single web page, and the relevant links are provided in that
page.  Every page follows the template corresponding to the kind of
concept it describes.  This homogeneity enables one to explore the
knowledge base confidently once the nature of the basic kinds of links
have been understood.

Thanks to the way the WWW is designed, the pages which constitute the
knowledge base need not all be in the same physical location. This
allows different parts to be developed by people working in different
locations. Of course, communication is necessary in order to establish
links between parts that are developed separately. The prototype is
accessible on the Web at the address
\texttt{http://www.iro.umontreal.ca/$\sim$paquetse/knoweb/000\_INTRODUCTION.html}.

This design has the following strengths:
\begin{itemize}
\item The architecture is simple and easy to understand.
\item Consulting the knowledge base is easy. All one needs is a WWW
browser and access to the Internet. Large numbers of people
nowadays have such access.
\item Contributing to the knowledge base is easy. All one needs
is the ability to write Web pages and means of making them
available. Large numbers of people nowadays have such means.
\item The architecture allows contributors to work together to
strengthen their respective contributions by linking them.
\item While restrictions exist on the format that is used to represent 
ideas, there is complete freedom in regard to the contents itself.
\end{itemize}

It shares the following shortcomings with the basic WWW architecture:
\begin{itemize}
\item Changing or refining existing parts of the knowledge base may
result in large numbers of links coming from other parts of the base
being invalidated. Work is then needed to fix those links.
\item In order to maximize their usefulness, new contributions must be properly linked 
to the existing base. This may require a lot of communication between
contributors.
\item There is no way of ascertaining the quality of contents that 
has been contributed.
\end{itemize}

We believe that despite the limitations of this initial design, such a 
concrete way of organizing ideas has the potential to win the favour
of dedicated individuals who not only like to learn but wish to
share their understanding with others in an efficient manner.

\section{Conclusion}
\label{conclusion}

We have proposed a simple methodology which aims at integrating
knowledge and know-how across disciplines in a coherent, usable
manner. This methodology is centered around
generalization/specialization relationships and a problem solving
perspective. We have provided a sample implementation of that
methodology which builds on the strengths of World Wide Web technology.

We hope that the reader feels the importance of the interdisciplinary
language barrier problem and sees this work as an interesting first
step towards solving it. He is currently actively seeking feedback and
help to improve this work. He can be reached by email at the address
\texttt{paquetse@iro.umontreal.ca}.

\bibliography{doc}

\end{document}